# Temperature-composition Phase Diagrams for $Ba_{1-x}Sr_xFe_2As_2$ & $Ba_{0.5}Sr_{0.5}(Fe_{1-y}Co_y)_2As_2$

Jonathan E. Mitchell,[1,*] Bayrammurad Saparov,[1] Wenzhi Lin,[2] Stuart Calder,[3] Qing Li,[2] Sergei V. Kalinin,[2] Minghu Pan,[2] Andrew D. Christianson,[3] and Athena S. Sefat[1,†]

[1]*Materials Science & Technology Division, Oak Ridge National Laboratory, Oak Ridge, TN 37831, USA*
[2]*Center for Nanophase Materials Sciences, Oak Ridge National Laboratory, Oak Ridge, TN 37831, USA*
[3]*Quantum Condensed Matter Division, Oak Ridge National Laboratory, Oak Ridge, TN 37831, USA*

Single crystals of mixed alkaline earth metal iron arsenide materials, $Ba_{1-x}Sr_xFe_2As_2$ with $0.185 \leq x \leq 0.762$ and $Ba_{0.5}Sr_{0.5}(Fe_{1-y}Co_y)_2As_2$ with $0.028 \leq y \leq 0.141$, are synthesized via the self-flux method. $Ba_{1-x}Sr_xFe_2As_2$ display spin-density wave features ($T_N$) at temperatures intermediate to the parent materials, $x = 0$ and 1, with $T_N(x)$ following an approximately linear trend. Cobalt doping of the 1 to 1 Ba:Sr mixture, $Ba_{0.5}Sr_{0.5}(Fe_{1-y}Co_y)_2As_2$, results in a superconducting dome with maximum transition temperature of $T_C = 19$ K at $y = 0.092$, close to the maximum transition temperatures observed in unmixed $A(Fe_{1-y}Co_y)_2As_2$; however, an annealed crystal with $y = 0.141$ showed a $T_C$ increase from 11 to 16 K with a decrease in Sommerfeld coefficient $\gamma$ from 2.58(2) to 0.63(2) mJ/(K$^2$ mol atom). For the underdoped $y = 0.053$, neutron diffraction results give evidence that $T_N$ and structural transition ($T_O$) are linked at 78 K, with anomalies observed in magnetization, resistivity and heat capacity data, while a superconducting transition at $T_C \approx 6$ K is seen in resistivity and heat capacity data. Scanning tunneling microscopy measurements for $y = 0.073$ give Dynes broadening factor $\Gamma = 1.15$ and a superconducting gap $\Delta = 2.37$ meV with evidence of surface inhomogeneity.

## I. INTRODUCTION

The discovery of superconductivity in F-doped LaFeAsO, with a $T_C$ of 26 K,[1] began a firestorm of research into a new class of layered iron-based high temperature superconductors (FeSC), which have given critical temperatures as high as 55 K.[2] One family of intensely studied FeSC materials is the so-called "122" family, with formula $AFe_2As_2$ ($A$ = Ba, Sr, Ca), which crystallize in the tetragonal $ThCr_2Si_2$-type structure ($I4/mmm$).[3-7] The $AFe_2As_2$ "parent" materials all display spin density wave (SDW) instability, with anomalies at $T_N \approx 140$ K, 205 K, and 170 K for $A$ = Ba, Sr, and Ca, respectively. These magnetic transitions are often accompanied by tetragonal-to-orthorhombic structural transitions near $T_N$.[7-9] Antiferromagnetic ordering can be suppressed to yield superconductivity through either chemical substitution,[10-20] or by applying hydrostatic pressure.[21-23]

Of particular interest has been the family of cobalt-doped 122s. In both Ba- and Sr-122, SDW behavior is quickly suppressed upon Co-doping, with a split between the antiferromagnetic transition and the tetragonal-to-orthorhombic structural transition ($T_O$) observed for the Ba-122 system,[9] although these transitions remain linked for Sr-122.[24] Superconductivity emerges above $y \sim 0.03$ for both Ba- and Sr-122.[9,25] $Ba(Fe_{1-y}Co_y)_2As_2$ displays a maximum $T_C$ of 22 K with $y \sim 0.08$,[18] while for $Sr(Fe_{1-y}Co_y)_2As_2$ the maximum $T_C$ of 19 K occurs at $y \sim 0.1$.[13] At Co-concentrations below the maximum $T_C$ superconductivity and antiferromagnetism coexist, as demonstrated from resistivity and magnetic susceptibility data[11] and supported by neutron scattering measurements.[26] For doping levels above the maximum $T_C$, the transition temperature tapers off until superconductivity disappears around $y \sim 0.18$, forming the typical dome-shaped superconducting region.

To date, there have been few studies of mixed alkaline earth metal 122s. Saha *et al* studied the crystallographic effects of substituting $Ba^{2+}$ and $Ca^{2+}$ into $SrFe_2As_2$, showing a trend of decreasing unit cell size as $A$ goes from Ba to Sr to Ca.[27] Recently, Kirshenbaum *et al.* found a relationship between the magnetic energy scale and the tetrahedral bond angles, which may be fine-tuned as a function of chemical pressure, the across the same $AFe_2As_2$ series.[28] Rotter *et al* compared single crystals of isovalently doped systems $Ba_{1-x}Sr_xFe_2As_2$ and $BaFe_2(As_{1-x}P_x)_2$, concluding that the Fe-As bond length was the primary driving force in suppressing SDW behavior to foment superconductivity.[29] Moreover, Wang *et al* used polycrystalline samples to demonstrate the effect of $Sr^{2+}$ content on both $T_N$ for $Ba_{1-x}Sr_xFe_2As_2$ as well as $T_C$ for $Ba_{1-x}Sr_x(Fe_{0.9}Co_{0.1})_2As_2$.[30]

Compared to these studies, we demonstrate the thermodynamic and transport properties in fine-tuned single crystals of $Ba_{1-x}Sr_xFe_2As_2$ ($0.185 \leq x \leq 0.762$) and $Ba_{0.5}Sr_{0.5}(Fe_{1-y}Co_y)_2As_2$ ($0 \leq y \leq 0.141$), grown using the self-flux method. Changes of structural lattice parameters are investigated using high-resolution powder x-ray diffraction, while the antiferromagnetic transition temperatures and the superconducting behavior are elucidated through a combination of magnetic susceptibility, resistivity, heat capacity, neutron diffraction, and scanning tunneling microscopy (STM) results.

## II. EXPERIMENTAL METHODS

Large single crystals ($\sim 8\times6\times0.2$ mm$^3$) of each material were grown from transition metal arsenide ($T$As) flux using high purity (>99.9 %) starting materials obtained from Alfa Aesar. FeAs and CoAs were pre-synthesized by heating evacuated silica ampoules containing unbound elements to 600 °C and holding for several hours. FeAs was further melted at 1065 °C before cooling. Appropriate stoichiometric quantities of both the alkaline earth metals ($AE$) and $T$As were mixed in alumina crucibles in a 1:5 ratio, sealed in silica tubes with a partial pressure of argon, heated to 1180 °C and held for a day. Each assembly was then cooled to 1090 °C at a rate of 2 °C/h, whereupon they were centrifuged to draw off the molten flux.

A few small crystals were ground for powder x-ray diffraction measurements, which were performed using a PANalytical X'Pert PRO MPD instrument fitted with a Ni-filtered Cu-K$_\alpha$ radiation source. Structural data was extracted via Rietveld refinement using GSAS-EXPGUI,[31,32] with all materials well indexed to the ThCr$_2$Si$_2$-type crystal structure with space group $I4/mmm$. Chemical composition of the materials was determined using a Hitachi TM3000 scanning electron microscope equipped with an energy-dispersive x-ray spectrometer (EDS). The atomic compositions of each batch were determined by averaging the EDS results from two spots on each of three separate crystals.

DC magnetization measurements were performed by a Quantum Design Magnetic Properties Measurement System (MPMS). Samples were initially cooled in the absence of a magnetic field (zero field cooled, ZFC) and data collected upon warming from 5 to 300 K with an applied field (1 T for non-superconducting materials, 20 Oe for superconductors) then cooled in the same field back to 5 K (field cooled, FC). Temperature dependent resistance, $R$(T), and heat capacity, $C_p$(T), data were collected using a Quantum Design Physical Property Measurement System (PPMS). $R$(T) was measured in the *ab*-plane employing the four-point probe method with platinum leads attached via Epo-Tek H20E silver epoxy and data collection in the range 1.8 to 300 K. $C_p$(T) was measured below 200 K by means of the relaxation method. Neutron scattering measurements for $Ba_{0.5}Sr_{0.5}(Fe_{0.947}Co_{0.053})_2As_2$ were performed on beamline HB-1 at the High Flux Isotope Reactor (HFIR), Oak Ridge National Laboratory. Measurements were carried out using a wavelength of $\lambda = 2.41$ Å in elastic mode with using a single crystal oriented in the *hhl* scattering plane.

## III. RESULTS AND DISCUSSION

### A. $Ba_{1-x}Sr_xFe_2As_2$

The room temperature lattice parameters, $c/a$ ratio, and unit cell volumes are given in Table 1 for $Ba_{1-x}Sr_xFe_2As_2$. The nominal $x$ values are provided along with those derived using EDS, with standard deviations included. While the strontium concentrations of the compounds with higher $x$ content lie within experimental error of the nominal value, there was a significant deviation for the nominally doped $x = 0.25$ specimen. The EDS derived $x$ values will be used to refer to $Ba_{1-x}Sr_xFe_2As_2$ materials throughout this work.

TABLE 1. Lattice parameters, $c/a$ ratio, and unit cell volumes for $Ba_{1-x}Sr_xFe_2As_2$. Both the nominal and EDS measured $x$ values are listed.

| $x$ (nominal) | $x$ (EDS) | $a$ (Å) | $c$ (Å) | $c/a$ | V (Å$^3$) |
|---|---|---|---|---|---|
| 0.25 | 0.185(10) | 3.95565(8) | 12.9113(4) | 3.2640 | 202.025(6) |
| 0.50 | 0.477(11) | 3.94448(8) | 12.7160(4) | 3.2237 | 197.847(6) |
| 0.75 | 0.762(4) | 3.93590(8) | 12.4897(4) | 3.1733 | 193.482(6) |

Figure 1 (a) shows a typical Rietveld refinement of the room temperature diffraction pattern for $Ba_{0.523}Sr_{0.477}Fe_2As_2$. Refinement indices for all three diffraction patterns were low, with profile factor $R_p = 1.85 - 2.34$ %, weighted profile factor $R_{wp} = 2.41 - 3.15$ %, and goodness of fit $\chi^2 = 1.14 - 1.37$. Unlike previously published data based on polycrystalline samples,[30] no peak broadening was observed for any of the ground crystals. Figure 1 (b) shows the trends in room temperature lattice dimensions $a$ and $c$, $c/a$ ratio, and unit cell volume with change in Sr content ($x$). In all cases, the values decrease monotonically with increasing $x$, due to the size discrepancy between $Ba^{2+}$ ($r_{ionic} = 1.42$ Å) and $Sr^{2+}$ ($r_{ionic} = 1.26$ Å)[34] cations. The trend matches well with those found in previous studies of mixed alkaline earth metal 122s.[27-30]

The ZFC temperature dependence of the magnetic susceptibility data are shown in Figure 2 (a). The SDW anomalies were determined as the peak maximum of the Fisher heat capacities, calculated as $d(\chi T)/dT$, giving $T_N$ values of 144 K, 166 K, and 182 K for $x = 0.185$, 0.477, and 0.762, respectively. Figure 2 (b) shows the temperature dependence of the resistivity normalized to 300 K for the $Ba_{1-x}Sr_xFe_2As_2$ series. The curves feature sharp downturns, in contrast to the broad transitions in resistivity observed by Wang *et al* from polycrystalline samples.[30] The transition temperatures derived from dρ/dT give 142 K, 160 K, and 179 K, for $x$ = 0.185, 0.477, and 0.762, respectively. These values agree with those determined from susceptibility measurements, and form an approximately linear trend between those reported for single crystals of BaFe$_2$As$_2$ ($T_N$ ~ 132 K)[33] and SrFe$_2$As$_2$ ($T_N$ ~ 203 K)[23]. In addition to the major features, there is also a small downturn at ~ 15 K for $x$ = 0.185, which may be the result of strain, nanoscale phase segregation, or have another electronic origin.

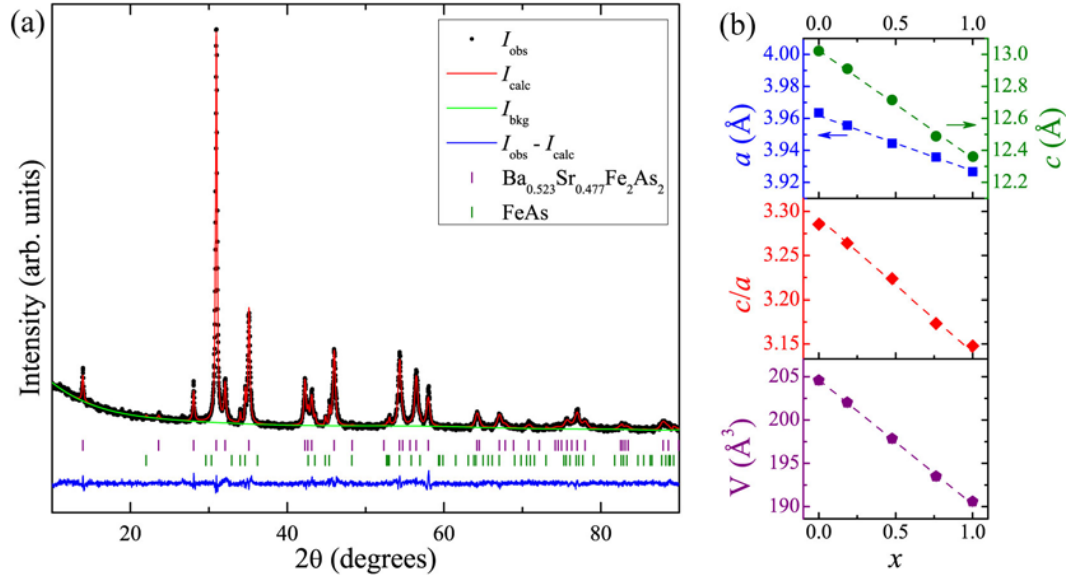

FIG. 1. (Color online) (a) Powder x-ray diffraction pattern (• symbols) for Ba$_{0.523}$Sr$_{0.477}$Fe$_2$As$_2$ with structural refinement via the Rietveld method (red line). Bragg reflection peaks are marked for both the main $I4/mmm$ phase (purple) and FeAs (green), due to residual surface flux. (b) Trends in lattice parameters, $c/a$ ratio, and unit cell volume with increasing $x$ for Ba$_{1-x}$Sr$_x$Fe$_2$As$_2$. Data for end members are taken from Refs. 33 and 23, respectively. Dashed lines are intended as a guide to the eye.

Heat capacity results are shown in Figure 3. Large peaks are evident with onsets at $T_N$ = 147 K, 165 K, and 182 K for $x$ = 0.185, 0.477, and 0.762, respectively, corresponding closely to the transitions obtained from the susceptibility and resistivity. The inset shows the fits of $C_p/T$ vs. $T^2$ at low temperature, which yield Sommerfeld electronic coefficients of $\gamma$ = 0.81(5) to 1.00(2) mJ/(K$^2$ mol atom).

Figure 4 shows the trend in $T_N$ with increasing Sr content for Ba$_{1-x}$Sr$_x$Fe$_2$As$_2$, with values derived from d($\chi$T)/dT, dρ/dT, and $C_p$ measurements. There is a linear increase in $T_N$ with increasing $x$, similar to the trend reported by Rotter et al[29] but in contrast to the non-linear trend found by Wang et al.[30] As noted above and shown in Table 1, the nominal versus experimental composition of alkaline earth metals in Ba$_{1-x}$Sr$_x$Fe$_2$As$_2$ may vary considerably, an issue unaccounted for in the latter publication.

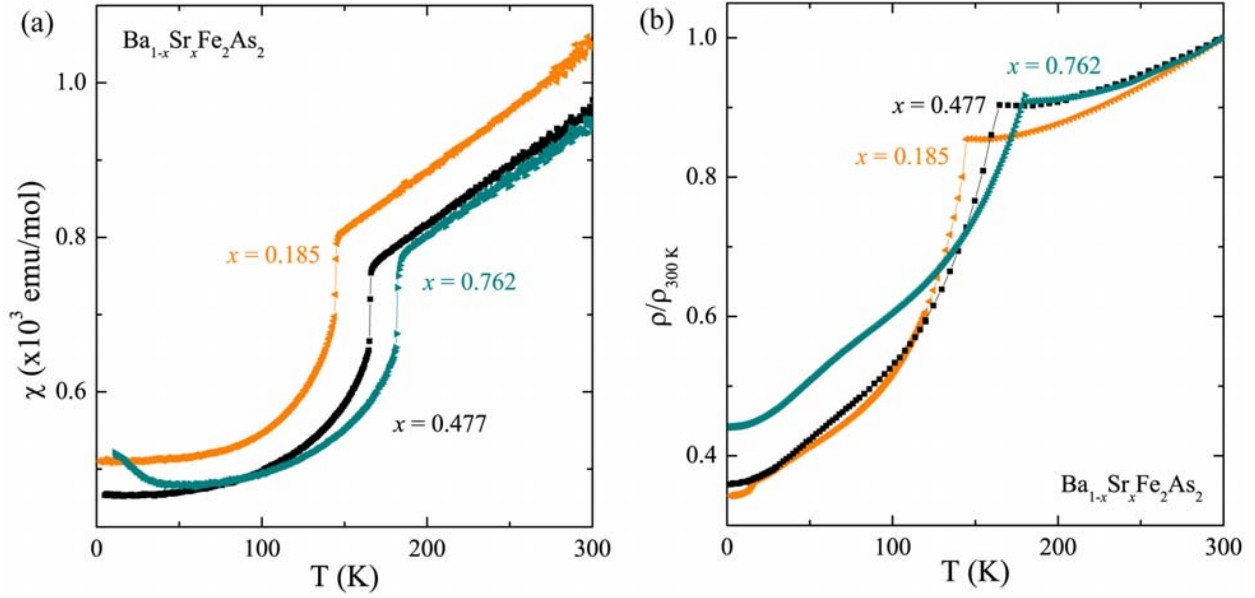

FIG. 2. (Color online) Temperature dependence of (a) magnetic susceptibility, and (b) normalized resistivity for $Ba_{1-x}Sr_xFe_2As_2$.

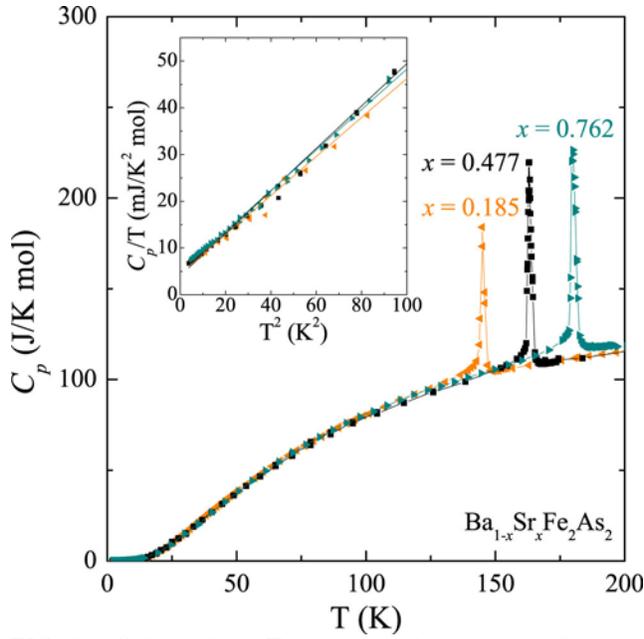

FIG. 3. (Color online) Temperature dependence of heat capacity, $C_p$, for $Ba_{1-x}Sr_xFe_2As_2$. Inset: $C_p/T$ vs. $T^2$ below 10 K. Solid lines represent fits linear fits to estimate Sommerfeld coefficient, $\gamma$.

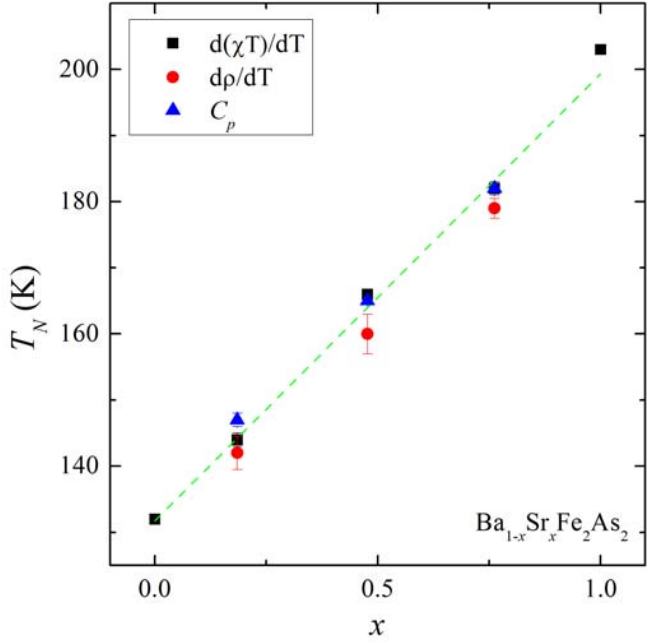

FIG. 4. (Color online) Trend in SDW transition temperatures with $x$ for $Ba_{1-x}Sr_xFe_2As_2$. The dashed green line is a linear fit of all data points. The values for end members $BaFe_2As_2$ and $SrFe_2As_2$ are from previous publications.[33,23]

## B. $Ba_{0.5}Sr_{0.5}(Fe_{1-y}Co_y)_2As_2$

Table 2 shows the nominal and experimentally determined $y$ values, as well as the Sr content, lattice parameters, $c/a$ ratio, and unit cell volume for the series $Ba_{0.5}Sr_{0.5}(Fe_{1-y}Co_y)_2As_2$, $0.028 \leq y \leq 0.141$. As with the non-Co-doped material described in section A, the EDS-verified strontium content is consistently a few percent lower than the nominal level of $x = 0.5$. Similarly, the experimental cobalt concentration ranges from 0.022 to 0.034 below the nominal doping level, with this gap widening as $y$ increases. For the purpose of simplicity, all materials will be referred to using the measured cobalt content, along with the nominal strontium content.

TABLE 2. Lattice parameters, $c/a$ ratio, and unit cell volumes for $Ba_{0.5}Sr_{0.5}(Fe_{1-y}Co_y)_2As_2$. The nominal $y$ values are provided along with the EDS-determined values for both $y$ and Sr content (nominally 0.5).

| $y$ (nominal) | $y$ (EDS) | Sr (EDS) | $a$ (Å) | $c$ (Å) | $c/a$ | V (Å$^3$) |
|---|---|---|---|---|---|---|
| 0.05 | 0.028(3) | 0.456(5) | 3.94557(5) | 12.6947(3) | 3.2175 | 197.624(4) |
| 0.075 | 0.053(3) | 0.469(15) | 3.94498(7) | 12.6785(3) | 3.2138 | 197.313(5) |
| 0.10 | 0.073(4) | 0.453(5) | 3.94555(5) | 12.6704(2) | 3.2113 | 197.245(4) |
| 0.125 | 0.092(2) | 0.460(6) | 3.94454(7) | 12.6585(3) | 3.2091 | 196.959(5) |
| 0.15 | 0.117(3) | 0.461(5) | 3.94547(5) | 12.6474(2) | 3.2055 | 196.878(4) |
| 0.175 | 0.141(2) | 0.468(11) | 3.94413(8) | 12.6277(4) | 3.2016 | 196.438(6) |

The powder x-ray diffraction pattern for highest $T_C$ $Ba_{0.5}Sr_{0.5}(Fe_{0.908}Co_{0.092})_2As_2$ is shown in Figure 5 (a) as representative for the Co-doped materials. Refinement indices for the diffraction patterns covered the ranges $R_p$ = 1.50 – 1.76 %, $R_{wp}$ = 1.90 – 2.25 %, and $\chi^2$ = 0.91 – 1.09. Figure 5 (b) shows the trends in lattice parameter values with respect to $y$. While the $a$- lattice parameter remains essentially constant as *y* increases, there is a contraction of the unit cell in the *c*-direction over the same interval, resulting in the decrease of the *c/a* ratio and unit cell volume.

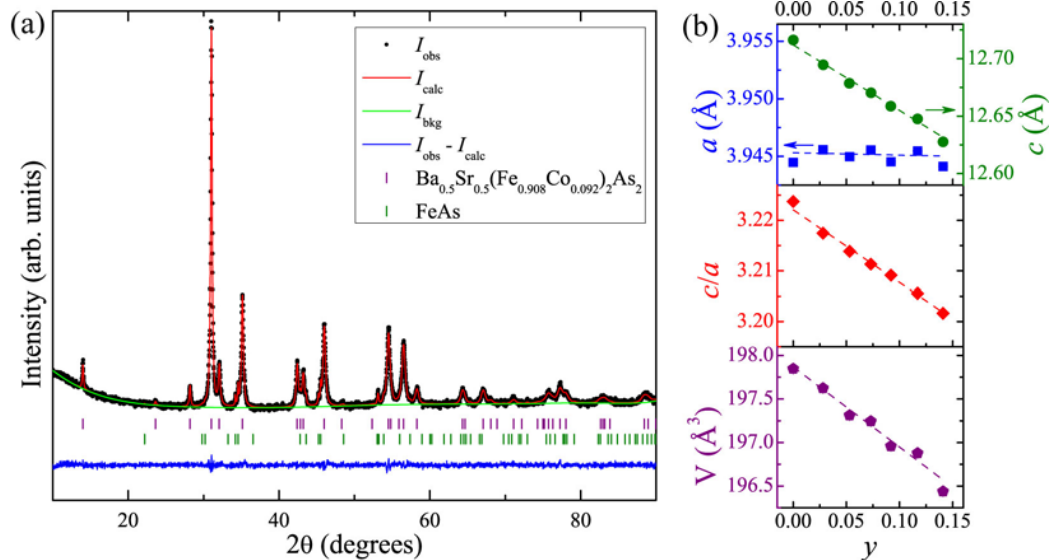

FIG. 5. (Color online) (a) Powder x-ray diffraction pattern (• symbols) for $Ba_{0.5}Sr_{0.5}(Fe_{0.908}Co_{0.092})_2As_2$ with structural refinement via the Rietveld method (red line). Bragg reflection peaks are marked for both the main *I4/mmm* phase (purple) and FeAs (green), due to residual surface flux. (b) Trends in lattice parameters, *c/a* ratio, and unit cell volume with increasing *y* for $Ba_{0.5}Sr_{0.5}(Fe_{1-y}Co_y)_2As_2$.

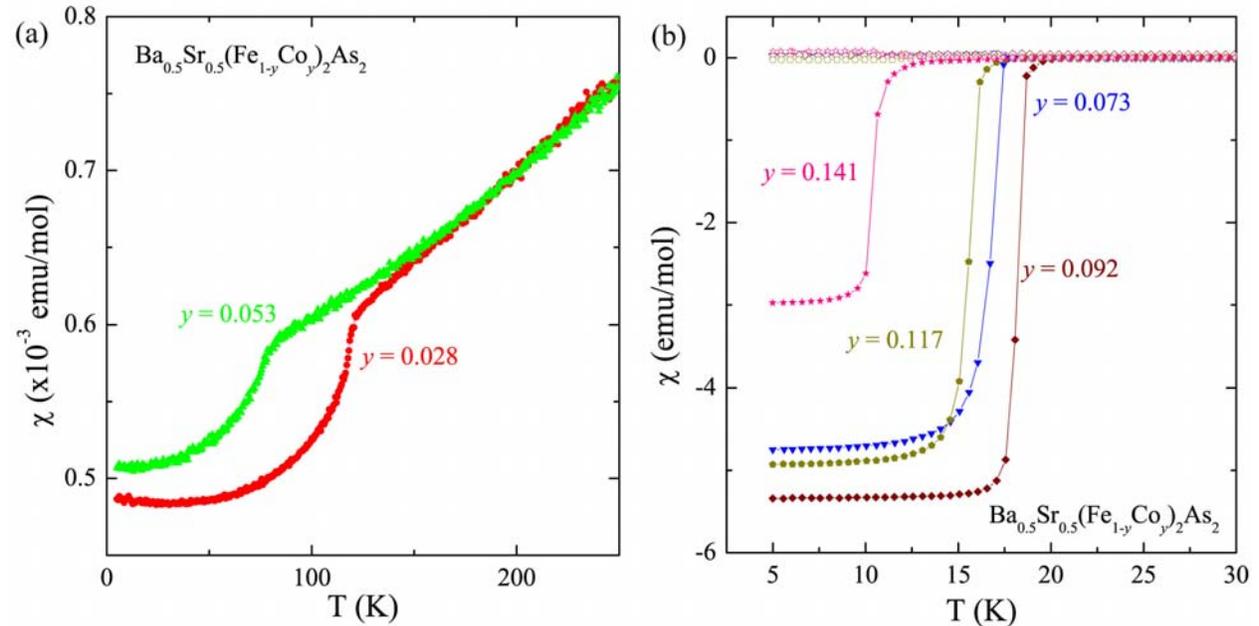

FIG. 6. (Color online) Temperature dependence of the magnetic susceptibility for $Ba_{0.5}Sr_{0.5}(Fe_{1-y}Co_y)_2As_2$. Behavior changes from SDW antiferromagnetism (a) to superconductivity (b) as Co-doping increases.

The temperature dependence of the magnetic susceptibility is shown in Figure 6. At low Co-doping levels, SDW rapidly decreases to $T_N = 78$ K for $y = 0.053$. Above this Co concentration superconductivity sets in, reaching a maximum $T_C$ (onset) = 19 K for $y = 0.117$, found by taking point of intersection of the tangents above and below the Meissner downturn. This $T_C$ value is unexpectedly high for a material with chemical disorder within both the superconducting $T$As ($T$ = Fe, Co) and the alkaline earth metal layers.

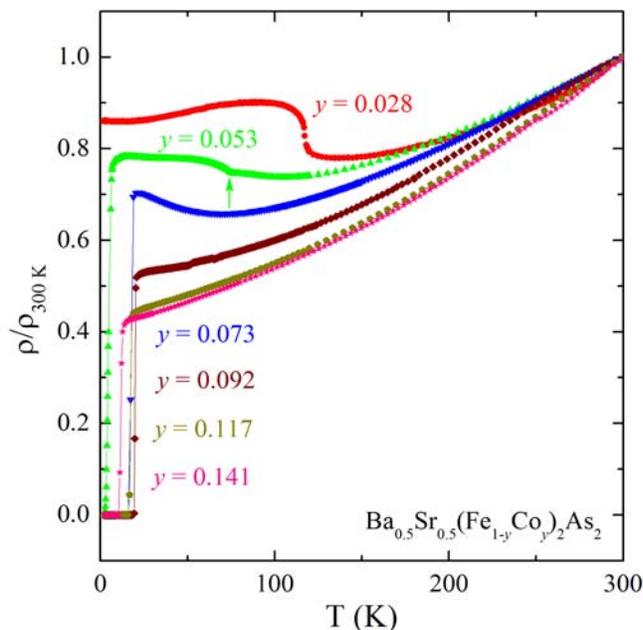

FIG. 7. (Color online) Normalized temperature dependence of resistivity for $Ba_{0.5}Sr_{0.5}(Fe_{1-y}Co_y)_2As_2$. The arrow points to the onset of antiferromagnetic ordering for $y = 0.053$.

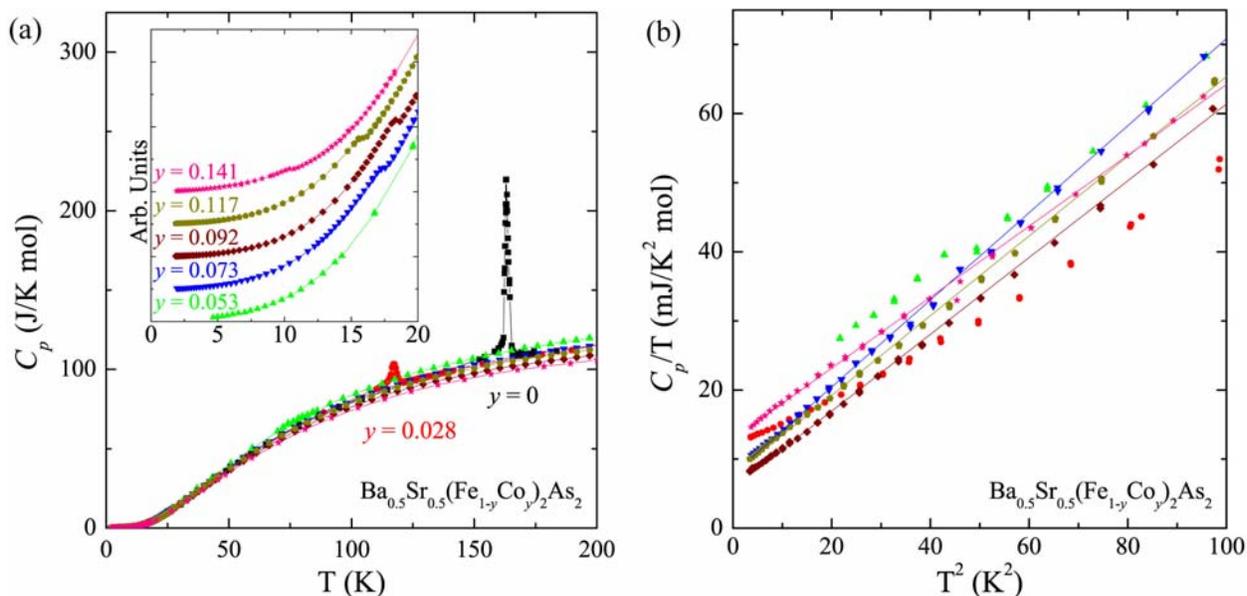

FIG. 8. (Color online) (a) Temperature dependence of heat capacity for $Ba_{0.5}Sr_{0.5}(Fe_{1-y}Co_y)_2As_2$. Inset: Data below 20 K for selected materials, stacked to emphasize weak anomalies due to superconductivity. (b) $C_p/T$ vs. $T^2$ below 10 K. Solid lines represent fits to estimate Sommerfeld coefficient, $\gamma$.

Figure 7 plots resistivity vs. temperature for $Ba_{0.5}Sr_{0.5}(Fe_{1-y}Co_y)_2As_2$ normalized to 300 K. A sudden increase in resistivity is observed for $y = 0.028$ around 117 K, corresponding to the SDW transition in the susceptibility data. All materials with higher Co-doping levels display a sharp drop to zero resistivity at low temperatures. For $y = 0.053$ the presence of antiferromagnetism alongside superconductivity is suggested by a slight increase in the resistivity, indicated by the green arrow. The derivative of the resistivity, $d\rho/dT$, results in a large peak at the superconducting transition $T_C = 5$ K, with a smaller peak at ~74 K, closely matching $T_N = 78$ K found in the susceptibility data. Similar features for single crystals of underdoped Ba-122 and Sr-122 have been attributed to the coexistence of competing antiferromagnetism and superconductivity.[11,26]

The heat capacity data for $Ba_{0.5}Sr_{0.5}(Fe_{1-y}Co_y)_2As_2$ are shown in Figure 8 (a). The peak at 117 K for $y = 0.028$ has a significantly diminished intensity compared to the non Co-doped samples ($y = 0$ is shown for comparison). All other materials show very small anomalies at low temperatures corresponding to superconducting transition temperatures obtained from resistivity data, although the peak for $y = 0.053$ is more easily discernable in the $C_p/T$ vs. $T^2$ plot displayed in Figure 8(b). The latter plot shows the trend below 10 K, which is approximately linear for all materials except $y = 0.028$. The $\gamma$ values range between 1.17(2) and 2.58(2) mJ/($K^2$ mol atom).

A few crystals of the $y = 0.141$ sample were sealed in evacuated quartz tubes and annealed at 800 ºC for one week. Magnetic, resistivity and heat capacity results show a 5 K increase in critical temperature to $T_C \sim 16$ K. In addition, the Sommerfeld coefficient decreased significantly from 2.58(2) to 0.63(2) mJ/($K^2$ mol atom). These effects are similar to what has been observed in previous annealing studies in the 122 family.[35,36] Whether the improvement in physical properties is a result of increased homogeneity of the Fe and Co distribution within the $T$As planes or of Ba and Sr in the spacer layers is not yet certain and is currently under investigation.

Figure 9(a) shows the emergence of intensity at the (0.5, 0.5, 3) reflection in neutron scattering measurements between 50 K and 90 K for $y = 0.053$. The observance of this peak indicates the existence of long-range magnetic order. Similar behavior was seen at the (0.5, 0.5, 1) reflection (not shown). The scattering intensity at (0.5, 0.5, 3) was measured through the magnetic transition between 55 K to 90 K (Figure 9(b)). Fitting the data to a power law yields a magnetic ordering temperature $T_N = 77.5(1.5)$ K, consistent with the anomalies from the bulk magnetic susceptibility data. Combining this with the resistivity and heat capacity data, a picture emerges of a material where antiferromagnetic ordering and superconductivity may coexist in this material, similar to what is seen upon Co-doping for both Ba- and Sr-122.[11,26] This trend is quickly suppressed as Co-content increases, with no observation of magnetism in $y = 0.073$.

The tetragonal-to-orthorhombic structural transition temperature ($T_O$) was also determined in the same neutron experiment. Although triple-axis measurements do not have sufficient resolution to directly observe peak splitting through the structural evolution, the transition is accompanied by a significant change in extinction, manifested by an observable difference in scattering intensity. The intensity of the nuclear (1,1,2) tetragonal reflection was followed through the same temperature regime studied for the magnetic reflection, showing an abrupt increase in intensity below the tetragonal-to-orthorhombic transition (Figure 9(c)). The inflection point indicates $T_O = 78.0(1.5)$ K, consistent with $T_N$. This shows that the behavior of $Ba_{0.5}Sr_{0.5}(Fe_{1-y}Co_y)_2As_2$ more closely follows that of $Sr(Fe_{1-y}Co_y)_2As_2$, where the magnetic and structural transitions occur at the same temperature,[24] rather than $Ba(Fe_{1-y}Co_y)_2As_2$ where antiferromagnetism sets in several kelvins below the structural transiton.

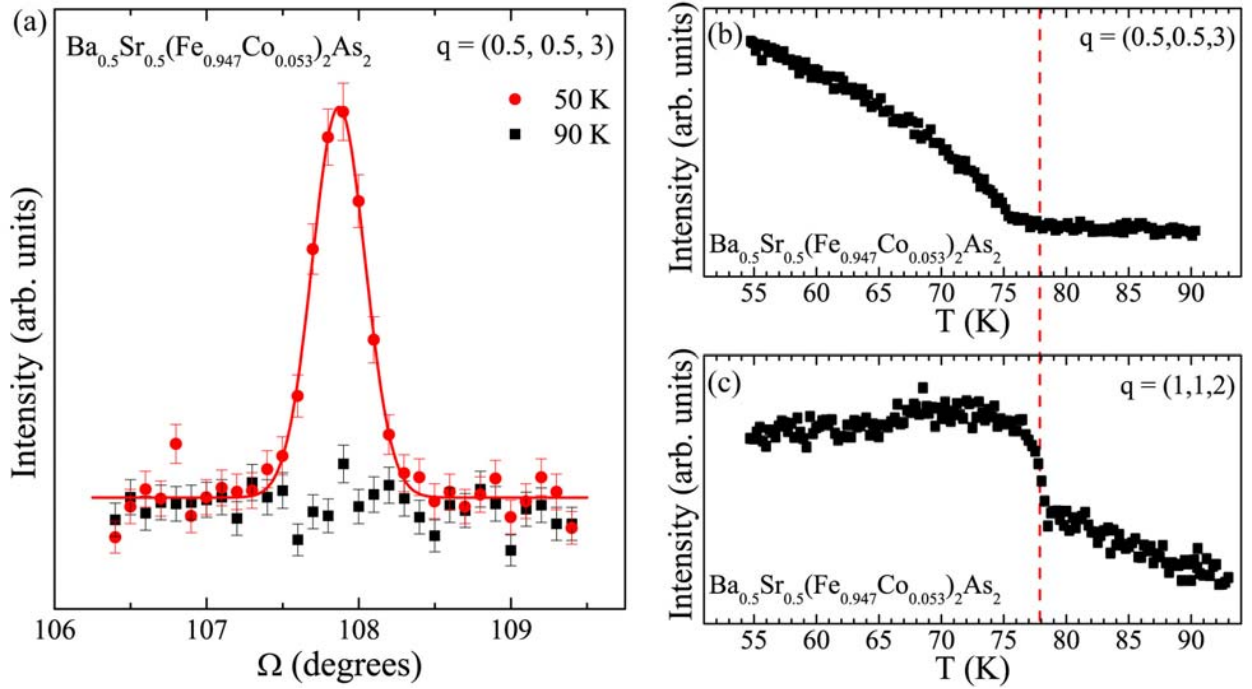

FIG. 9. (Color online) (a) Neutron scattering rocking scans for $Ba_{0.5}Sr_{0.5}(Fe_{0.947}Co_{0.053})_2As_2$ around the q = (0.5, 0.5, 3) magnetic reflection at $T$ = 50 K and 90 K, showing the development of magnetic ordering. (b) Temperature dependence of the integrated intensity of the scattering at q = (0.5, 0.5, 3). The curve is a fit to a power law indicating $T_N$ = 77.5(1.5) K. (c) Temperature dependence of the integrated intensity of the scattering at q = (1, 1, 2). The inflection point of the curve indicates the structural transition $T_O$ = 78.0(1.5) K.

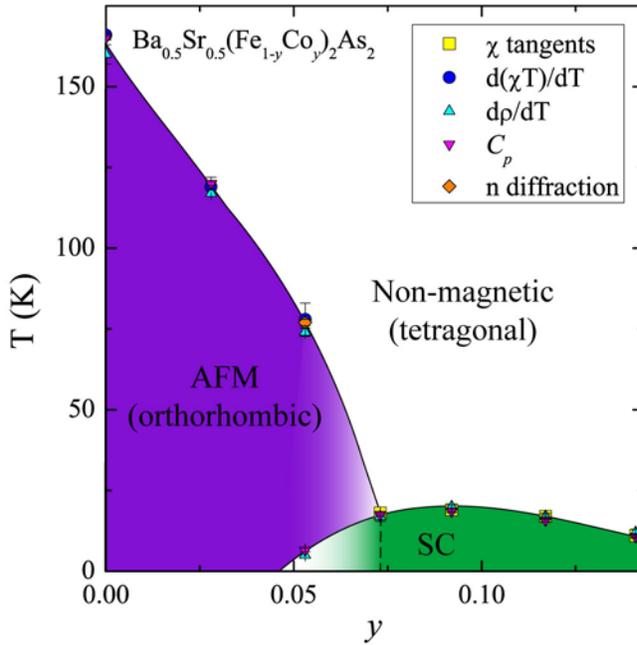

FIG. 10. (Color online) Phase diagram of transition temperature vs. $y$ for $Ba_{0.5}Sr_{0.5}(Fe_{1-y}Co_y)_2As_2$, showing the transition from non-magnetic to antiferromagnetic SDW behavior (AFM) and the dome of superconductivity (SC). A high-temperature tetragonal to low-temperature orthorhombic structural phase transition is expected, similar to what is seen for other transition-metal doped Ba/Sr-122 systems.[9,24]

A phase diagram may be constructed for $Ba_{0.5}Sr_{0.5}(Fe_{1-y}Co_y)_2As_2$ transition temperatures as a function of Co-concentration (Figure 10). This figure bears strong similarities between the phase diagrams of Co-doped Ba- and Sr-122, both of which show rapid suppression of SDW with $y$, a region with the possible coexistence of magnetism and superconductivity, and $T_C(max)$ around ~ 20 K with $y$ ~ 8 – 10 %.[9,13] The Ba/Sr-122 system shows features intermediate to the parent materials. Similar to Sr-122[24] but unlike Ba-122,[9,37] neutron experiments demonstrate that $T_O$ and $T_N$ transitions are coupled.

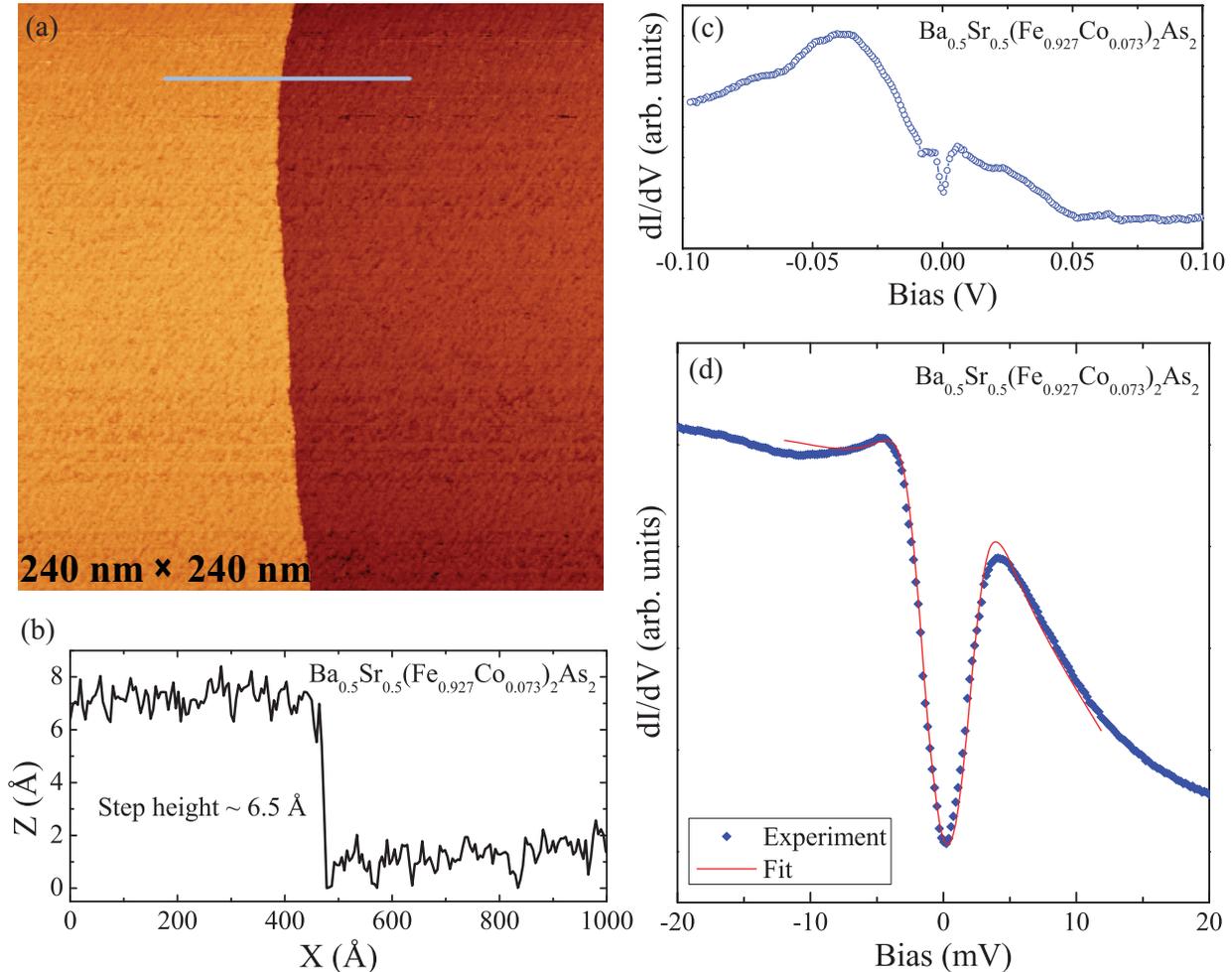

FIG. 11. (Color online) (a) Large scale topographic image of a single crystal of $Ba_{0.5}Sr_{0.5}(Fe_{0.927}Co_{0.073})_2As_2$ over a step. The setup conditions for imaging were a sample-bias voltage of +0.8 V and a tunneling current of 20 pA. (b) Line profile across the step, measured following the grey line in (a). (c) Differential tunneling conductance spectrum taken over a large energy range, with an overall background asymmetry. (d) Differential tunneling conductance spectrum data (dotted curve) with fitted curve. Data were taken with a sample-bias voltage of –50 mV and a tunneling current of 0.1 nA. Bias-modulation amplitude was set to 0.3 mV$_{rms}$.

Scanning tunneling microscopy (STM) and scanning tunneling spectroscopy (STS) were used to study the sample surface of the $y = 0.073$ sample of $Ba_{0.5}Sr_{0.5}(Fe_{1-x}Co_x)_2As_2$ after cleaving the crystals in ultrahigh vacuum. Figure 11(a) displays a sharp step in an overall flat surface in a large scale image taken at about 79 K. A line profile was measured across this step, indicated by the grey line. The measured step height in the line profile is ~ 6.5 Å, corresponding to a half-unit-cell step (Figure 11(b)). Doubling this height (~ 13 Å) provides a value which is in good agreement with out-of-plane lattice parameter $c = 12.6704(2)$ Å. STS data taken at 4.3 K demonstrates a clear superconducting gap, and an overall background asymmetry in the high energy range (Figure 11(c)). For the purpose of extracting the magnitude of the superconducting energy gap $\Delta$,[38] the STS spectrum shown in Figure 11(d) was fit according to the following procedure.[39] The background was first removed using polynomial functions to fit the background for the positive and negative bias regions and extend to Fermi energy. By the use of s-wave BCS gap function, the STS spectrum was subsequently fitted with a Dynes broadening factor $\Gamma$,[40] which is convoluted with the Fermi function at 4.2 K. This resulted in the fitting parameters $\Gamma = 1.15$ and $\Delta = 2.37$ meV, which yields the ratio $2\Delta/k_BT_C = 3.2$, with $T_C = 17$ K. The fitted curve is in good agreement with the STS data shown by the dotted curve (Figure 10(d)). We have not been able to determine the gap symmetry based on the above fitting due to the arbitrary nature of the background as well as the relatively high measurement temperature with respect to the superconducting critical temperature $T_C$ (17 K).

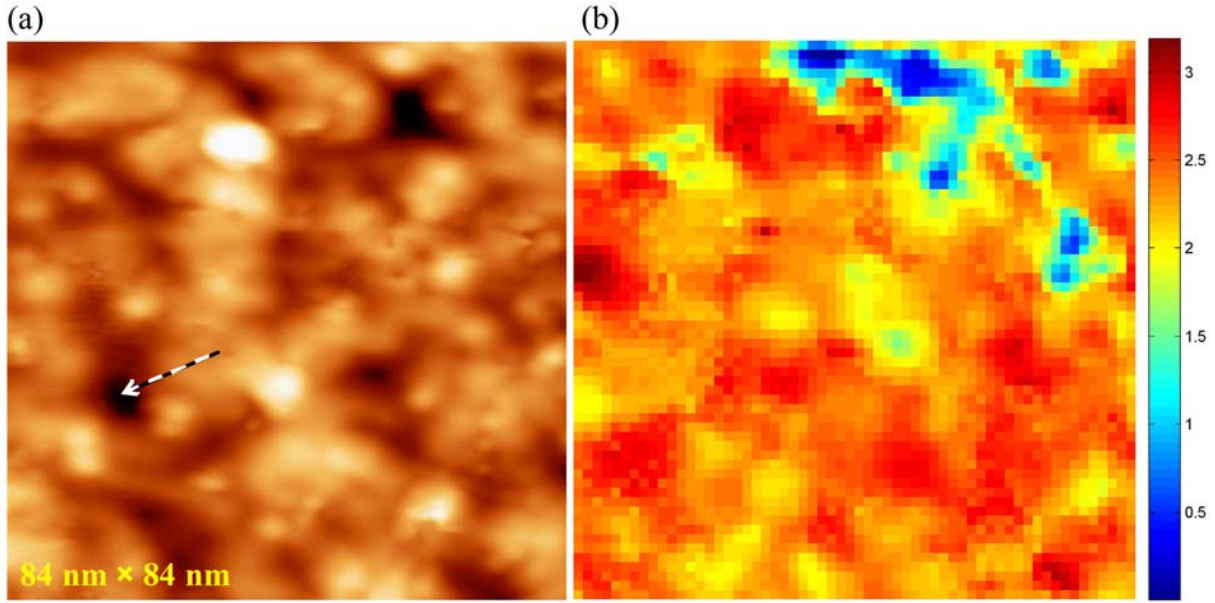

FIG. 12. (Color online) (a) Topographic image of a single crystal surface of $Ba_{0.5}Sr_{0.5}(Fe_{0.927}Co_{0.073})_2As_2$. The setup conditions for imaging were a sample-bias voltage of +0.05 V and a tunneling current of 200 pA. (b) Superconducting gap map yielded by analyzing differential tunneling conductance spectra taken over 60×60 grids in the same area shown in (a). Spectra were taken with a sample-bias voltage of +50 mV and a tunneling current of 0.2 nA. Bias-modulation amplitude was set to 0.5 mV$_{rms}$. The gap value is plotted as function of spatial location.

To address the spatial variation of superconducting gap value, spectrum surveys over 60×60 grids were taken in the same area as the topographic image in Figure 12(a), which was taken over an 84nm ×84 nm area at about 4.3 K. The image shows thin clusters as a common feature at the top of the surface. These clusters are well connected in many places but poorly connected in other places, as marked with the arrow. The grids contain differential tunneling conductance spectra (dI/dV versus V) at each spot of 60×60 grids over the area. By applying the fitting outlined above to each differential tunneling conductance spectrum, the STS measurements are analyzed to yield an energy gap map (Figure 12(b)), in which spatial variation of energy gap value $\Delta$ is displayed. In most of the area studied, the gap values are approximately 2.5 meV, although a large range from about 3 meV down to 0 meV is observed, indicating inhomogeneity of gap values over the surface.

## IV. CONCLUSION

The effects of mixing alkaline earth metals Ba and Sr in the 122 family of iron-based superconductors is studied. The lattice parameters in the $Ba_{1-x}Sr_xFe_2As_2$ series decrease monotonically with $x$, while SDW transition temperature increases over the same range. The $Ba_{0.5}Sr_{0.5}(Fe_{1-y}Co_y)_2As_2$ series shows a linear decrease in the $c$-lattice parameter with an essentially constant $a$-lattice parameter. Magnetic susceptibility, resistivity, heat capacity and neutron scattering data were used to construct temperature vs. concentration phase diagram. The Co-doped Ba/Sr-122 system shows features intermediate between that of the parent series, with rapid suppression of SDW transition temperature, a region where antiferromagnetism and superconductivity likely coexist, and a maximum $T_C \sim 19$ K at $y = 0.092$. This high superconducting transition temperature is surprising in a material with chemical disorder both within and without the superconducting FeAs planes. Although STM measurements confirm electronic inhomogeneity of the crystal surfaces, thermal annealing is found to improve $T_C$ by $\sim 5$ K in $y = 0.141$. STM and TEM measurements are currently under way to find evidence for micro- and nanostructural order within annealed crystals.

## ACKNOWLEDGEMENTS

Research was primarily supported by the U.S. Department of Energy, Basic Energy Sciences, Materials Sciences and Engineering Division. Part of this research was conducted at the Center for Nanophase Materials Sciences (CNMS) and at the High Flux Isotope Reactor (HFIR), which are sponsored at Oak Ridge National Laboratory by the Scientific User Facilities Division, Office of Basic Energy Sciences, U.S. Department of Energy.